\documentclass[conference]{IEEEtran}

\usepackage{booktabs}
\usepackage{siunitx}
\usepackage{color}
\usepackage{tabularx}
\usepackage{multirow}
\usepackage{balance}
\usepackage{scrextend}
\usepackage{listings}
\usepackage{dsfont}
\usepackage{subcaption}
\usepackage{tikz}
\usetikzlibrary{positioning}
\usepackage{pifont}
\usepackage{url,xspace}
\usepackage{algorithm}
\usepackage{amsmath}
\usepackage{bigstrut,cite}
\usepackage{float}
\usepackage[bookmarks=false]{hyperref}
\usepackage{cleveref}
\usepackage{bold-extra}
\usepackage{amsmath}
\usepackage{amsthm}
\usepackage{hhline}
\usepackage{adjustbox}
\newtheorem{definition}{Definition}

\newcommand{\EH}{EmptyHeaded}

\newtheorem{example}{Example}

\def\compactify{\itemsep=0pt \topsep=0pt \partopsep=0pt \parsep=0pt}

\begin{document}
%
\title{Old Techniques for New Join Algorithms: A Case Study in RDF Processing}

\author{\IEEEauthorblockN{Christopher R. Aberger}
\IEEEauthorblockA{caberger@stanford.edu}
\and
\IEEEauthorblockN{Susan Tu}
\IEEEauthorblockA{sctu@stanford.edu}
\and
\IEEEauthorblockN{Kunle Olukotun}
\IEEEauthorblockA{kunle@stanford.edu}
\and
\IEEEauthorblockN{Christopher R\'e}
\IEEEauthorblockA{chrismre@cs.stanford.edu}}

\maketitle

\begin{abstract}

Recently there has been significant interest around designing specialized RDF engines, as traditional query processing mechanisms incur orders of magnitude performance gaps on many RDF workloads. At the same time researchers have released new worst-case optimal join algorithms which can be asymptotically better than the join algorithms in traditional engines. In this paper we apply worst-case optimal join algorithms to a standard RDF workload, the LUBM benchmark, for the first time. We do so using two worst-case optimal engines: (1) LogicBlox, a commercial database engine, and (2) \EH, our prototype research engine with enhanced worst-case optimal join algorithms. We show that without any added optimizations both LogicBlox and \EH{} outperform two state-of-the-art specialized RDF engines, RDF-3X and TripleBit, by up to 6x on cyclic join queries---the queries where traditional optimizers are suboptimal. On the remaining, less complex queries in the LUBM benchmark, we show that three classic query optimization techniques enable \EH{} to compete with RDF engines, even when there is no asymptotic advantage to the worst-case optimal approach. We validate that our design has merit as \EH{} outperforms MonetDB by three orders of magnitude and LogicBlox by two orders of magnitude, while remaining within an order of magnitude of RDF-3X and TripleBit.

\end{abstract}

\IEEEpeerreviewmaketitle
\section{Introduction}

The volume of Resource Description Framework (RDF) data from the Semantic Web has grown exponentially in the past decade \cite{neumann2010rdf,yuan2013triplebit}. RDF data is a collection of Subject-Predicate-Object triples that form a complex and massive graph that traditional query mechanisms do not handle efficiently \cite{klyne2006resource,neumann2010rdf}. As a result, there has been significant interest in designing specialized engines for RDF processing \cite{neumann2010rdf,yuan2013triplebit,kim2015taming,atre2010matrix}. These specialized engines accept the SPARQL query language and build several indexes ($>10$) over the Subject-Predicate-Object triples to process RDF workloads efficiently \cite{neumann2010rdf,yuan2013triplebit}. In contrast, the natural way of storing RDF data in a traditional relational engine is to use triple tables \cite{neumann2010rdf} or vertically partitioned column stores \cite{abadi2007scalable}, but these techniques can be three orders of magnitude slower than specialized RDF engines \cite{neumann2010rdf}. The goal of this paper is to reexamine the performance difference between specialized RDF engines and general-purpose relational engines with a new class of worst-case optimal join algorithms.

Recent database theory has shown that Selinger-style join optimizers \cite{astrahan1976system}, used inside of both traditional relational engines and specialized RDF engines, can be asymptotically suboptimal due to their computation of joins in a pairwise fashion \cite{ngo2012worst}.  In fact, on cyclic queries, which can be the bottleneck queries in standard RDF benchmarks \cite{guo2005lubm}, any pairwise relational plan is provably worse asymptotically. For example, both RDF and traditional relational engines execute the ``triangle listing'' query, found in the standard LUBM RDF benchmark \cite{guo2005lubm}, in time $O(N^{3/2})$, where $N$ is the number of tuples in the input relation. Any traditional query plan takes $\Omega(N^2)$ and is thus suboptimal by a factor of $\sqrt{N}$. Fortunately, recent database theory has suggested new multi-way style join algorithms that solve arbitrary join patterns with a series of set intersections and loops to guarantee worst-case optimality~\cite{ngo2012worst}.

These new worst-case optimal join algorithms can support rich applications as shown by the LogicBlox engine~\cite{aref2015design} which is the first commercial database engine to use a worst-case optimal join algorithm. In fact, on the aforementioned cyclic queries, where worst-case optimal join algorithms have an asymptotic advantage, LogicBlox can be 18x faster than traditional relational engines and 4x faster than specialized RDF engines (see \Cref{sec:experiments}). Nevertheless, LogicBlox does not come with fully optimized query plans or indexes. Thus, LogicBlox can be two orders of magnitude slower than specialized engines on queries where they do not have an asymptotic advantage (see \Cref{sec:experiments}). This leaves open the question of whether worst-case optimal join algorithms are practically beneficial across RDF workloads--or whether there is a fundamental inefficiency in this style of engine for some RDF patterns. The key question we ask is {\it to what extent can these new worst-case optimal join algorithms achieve performance comparable to that of the specialized RDF engines?}

To answer this question, we use \EH{} \cite{aberger2015emptyheaded}, our prototype worst-case optimal engine designed around recent advancements in join processing \cite{joglekar2015aggregations}. \EH{} uses query compilation techniques and optimized data layouts that are both different from those in LogicBlox's original design.  To determine the performance gap between worst-case optimal engines and state-of-the-art specialized RDF engines, we compare LogicBlox and \EH{} against RDF-3X~\cite{neumann2010rdf} and TripleBit~\cite{yuan2013triplebit} on the entire LUBM benchmark~\cite{guo2005lubm}.

We find that by adding three classical (and simple) query processing optimizations to \EH{}, we are able to consistently achieve performance within an order of magnitude of specialized RDF engines, while sometimes outperforming them. The first optimization we consider is optimized index layouts for the worst-case optimal join algorithm. This optimization provides up to a 8x speedup on simple RDF patterns with high selectivity. The second optimization we consider is pushing down selections in our query plans. This can enable up to a 234x speedup on RDF patterns with high selectivity. Finally, our third optimization is pipelining intermediate results across nodes in our query plan. This can provide up to a 4x performance improvement. In general, we find these three classical optimizations to be necessary for worst-case optimal join algorithms to compete with specialized RDF engines.

Our contributions are as follows:
\begin{itemize}
\item We are the first to benchmark engines with a worst-case optimal join algorithm on a standard RDF workload. We show that on cyclic join queries, the bottleneck queries in the LUBM benchmark, that a worst-case optimal design outperforms both specialized and traditional data processing engines by up to 6x without any added optimizations.
\item We map three classic query optimizations to a worst-case optimal engine: (1) optimized index layouts, (2) pushing down selections, and (3) pipelining intermediate results. We show that these optimizations improve performance by over two orders of magnitude and allow \EH{} to become competitive with TripleBit and RDF-3X on simple acyclic queries with high selectivity.
\item We validate that \EH{} consistently outperforms traditional relational engines and existing worst-case optimal engines while remaining competitive with specialized RDF engines. Our performance can be two orders of magnitude better than the traditional relational design of MonetDB, an order of magnitude better than LogicBlox, and within an order of magnitude of TripleBit and RDF-3X.
\end{itemize}

We hope our work serves as a feasibility study, validating that worst-case optimal join algorithms have merit and can serve as an improvement over the traditional querying processing mechanisms used in common RDF workloads (in some cases more than others).

\section{Background}
\label{sec:backgound}
The EmptyHeaded engine works in three phases: (1) the query compiler translates a high-level datalog-like query into a logical query plan represented using a generalized hypertree decomposition (GHD)~\cite{green2007provenance}, replacing traditional relational algebra based query plans; (2) code is generated for the execution engine by translating the GHD into a series of set intersections and loops; and (3) the execution engine performs automatic algorithmic and layout decisions. 

For completeness, we summarize the design points of the \EH{} engine necessary to understand the added optimizations in \Cref{sec:optimizations}. We present these results informally and refer the reader to Aberger et al.~\cite{aberger2015emptyheaded} for a complete survey. We recapitulate the data representation, worst-case optimal join algorithm, and query plans used inside of \EH{}.

\subsection{Data Representation}

\begin{figure}
  \centering
  \includegraphics[width=0.95\linewidth]{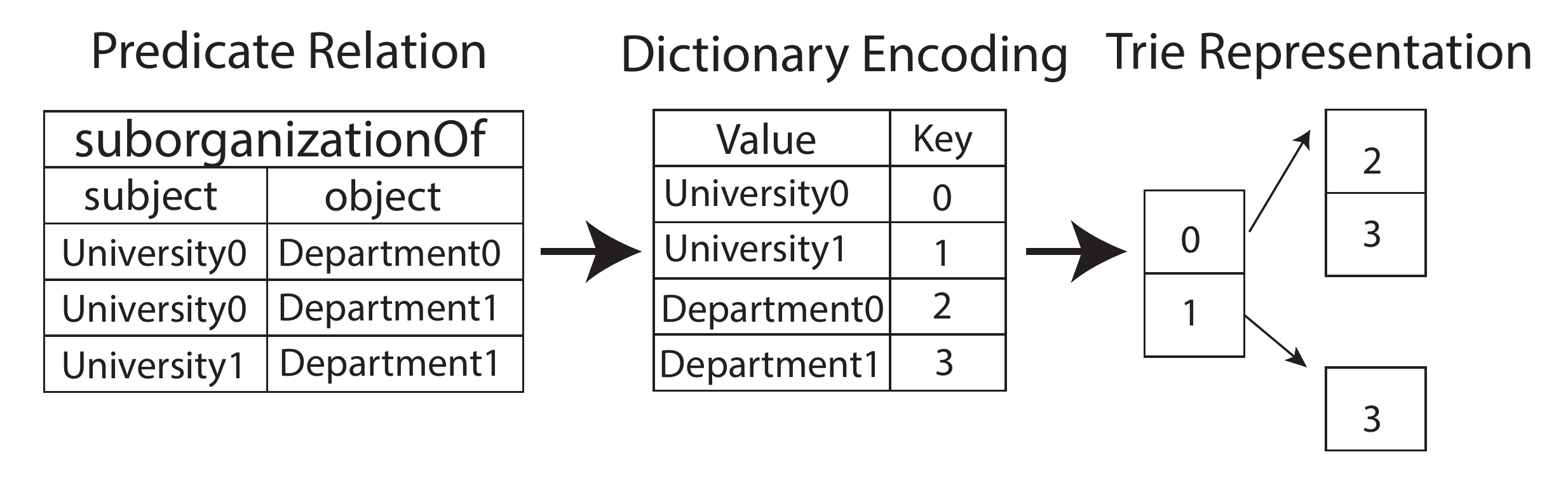}
  \caption{Transformation process from a vertically partitioned relation to \EH's trie representation.}
  \label{fig:trie}
\end{figure}

\EH{} stores all relations (input and output) using tries, which are multi-level data
structures common in column stores and graph engines \cite{stonebraker2005c,Hong:2012:GDE:2150976.2151013}.  The tries in \EH{} currently support sets containing 32-bit values stored using either (1) an unsigned integer array or (2) a bitset layout.

\subsubsection{Trie Structure}

Prior to building a trie, \EH{} performs dictionary encoding \cite{neumann2010rdf,yuan2013triplebit}, to encode relations of arbitrary types into 32-bit values. Dictionary encoding maps original data values to keys of another type---in our case 32-bit unsigned integers. After dictionary encoding, our 32-bit value relations are grouped into sets of distinct values based on a previous (if present) attribute or column. Each level of the trie corresponds to an attribute or column of an input relation. The mapping of levels to attributes is not fixed. We choose the order of the attributes for the trie based on a global attribute order, which is analogous to selecting a single index over the relation. We describe how we select a global attribute order in \Cref{sec:query_plans} and show the complete transformation from a vertically partitioned table \cite{abadi2007scalable} to a trie in \Cref{fig:trie}.

\subsubsection{Set Layouts}
\EH{} uses an automatic layout optimizer to select data layouts for each set within the trie data structure. Currently, \EH{} chooses between an unsigned integer array and bitset layout. \EH{} has a set optimizer that chooses the layout for each set in isolation based on its cardinality and range. The optimizer chooses the bitset layout when more than one out of every 256 values appears in the set.\footnote{The size of an AVX register.} It otherwise defaults to the unsigned integer array layout. We found that these layout decisions provided over an order of magnitude performance improvement on join patterns where set intersection is the bottleneck operation \cite{aberger2015emptyheaded}.

\subsection{Worst-Case Optimal Joins}

The generic worst-case optimal join algorithm serves as the foundation for the join algorithms used in the LogicBlox and \EH{} engines. The generic worst-case optimal join can be asymptotically better than any pairwise join plan. As an example, consider query 2 in the LUBM benchmark \cite{guo2005lubm}:
{
\footnotesize
\begin{gather*}
\texttt{undergraduateDegreeFrom(x,y)} \bowtie \texttt{memberOf(x,z)} \\
\bowtie \texttt{subOrganizationOf(z,y)} \\ \bowtie \texttt{type(x,a=`GraduateStudent')} \\
\bowtie \texttt{type(y,b=`University')} \\
\bowtie \texttt{type(z,c=`Department')}
\end{gather*}
}

\begin{algorithm}
  \begin{lstlisting}[
        basicstyle = \footnotesize,
        language = C,
        numbers = left,
        framexleftmargin=1.5em,
        xleftmargin=2em,
        mathescape]
  //Input: Hypergraph $H = (V,E)$, and a tuple $t$.
  Generic-Join($V$,$E$,$t$):
    if $|V| = 1$ then return $\cap_{e \in E} R_e[t]$.
    $I \leftarrow \{ v_{1} \}$ // the first attribute in V.
    Q $\leftarrow \emptyset$ // the return value
    // Intersect all relations that contain $v_{1}$
    // Only those tuples that agree with $t$.
    for every $t_v \in \cap_{e \in E : e \ni v_{1}} \pi_{I} (R_e[t])$ do
      $Q_t$ $\leftarrow$ Generic-Join($V - I$, $E$, $t :: t_v$ )
      Q $\leftarrow Q \cup \{ t_v \} \times Q_t $
    return Q
  \end{lstlisting}
  \caption{Generic Worst-Case Optimal Join Algorithm}
  \label{fig:worst_case}
\end{algorithm}

This query contains the following cyclic subquery which forms a triangle pattern:
{
\footnotesize
\begin{gather*}
\texttt{undergraduateDegreeFrom(x,y)} \bowtie \texttt{memberOf(x,z)} \\
\bowtie \texttt{subOrganizationOf(z,y)}
\end{gather*}
}
On this cyclic subquery, the generic worst-case optimal join algorithm runs in a time proportional to the maximum number of tuples
that could be output. Assuming the three relations are all of size $N$, the worst-case output size is $O(N^{3/2})$ here, whereas, any pairwise plan has a worst-case runtime $\Omega(N^2)$. The generic worst-case optimal join algorithm is presented in \Cref{fig:worst_case}.

In fact, for {\em any} join query, the generic worst-case optimal join algorithm's execution time can be upper bounded by the Atserias, Grohe, and Marx (AGM) bound \cite{agm}. This can be easily computed when the query is represented as a hypergraph. A \textbf{hypergraph} is a pair $H = (V, E)$, consisting of a nonempty set $V$ of vertices, and a set $E$ of subsets of $V$, the hyperedges of $H$. There is a vertex for each attribute of the query and a hyperedge for each relation. Now, fix a hypergraph $H$.  The AGM bound tells us that the output size is upper bounded by $$\prod_{e \in E} |R_e|^{x_e}$$ under the constraints

$$\forall v \in V, \sum_{e\in E: e \ni v} x_e \geq 1$$ $$\forall e \in E, x_e \geq 0$$
To get the tightest bound, we would like to minimize $\prod_{e \in E} |R_e|^{x_e}$, subject to those constraints. 

\subsection{Query Plans}
\label{sec:query_plans}

\EH{} uses GHDs to represent the query plans. GHDs enable optimizations such as pushing down selections, pushing down projections, and early aggregation that engines based solely on the generic worst-case optimal join algorithm are unable to capture. We briefly summarize how GHDs are used in \EH{}. For a more detailed discussion we refer the reader to Aberger et al. \cite{aberger2015emptyheaded}.

\begin{definition}
 Let $H$ be a hypergraph. A {\bf {\em generalized hypertree decomposition
    (GHD)}} of $H$ is a triple $D=(T, \chi, \lambda)$, where:
\begin{itemize}\compactify
\item  $T(V(T), E(T))$ is a tree
\item $\chi: V (T) \rightarrow 2^{V(H)}$ is a function associating a set of vertices $\chi(t) \subseteq V(H)$ to each node $t$ of $T$
\item $\lambda: V(T) \rightarrow 2^{E(H)}$ is a function associating a set of hyperedges to each node $t$ of $T$
\end{itemize}

such that the following properties hold:
\begin{itemize}
\item[1.] For each $e \in E(H)$, there is a node $t \in V(T)$ such that $e \subseteq \chi(t)$. 
\item[2.] For each $v \in V(H)$, the set $\{t \in V(T) | v \in \chi(t)\}$ is connected in $T$.
\item[3.] For every $t \in V(T)$, $\chi(t) \subseteq \cup
  \lambda(t)$.
\item[4. ] For every $t \in V(T)$, $\chi(t) \subseteq \cup_{e \in \lambda(t)} e$
\end{itemize}

\end{definition}

\begin{figure}
  \centering
  \includegraphics[width=0.8\linewidth,height=4cm]{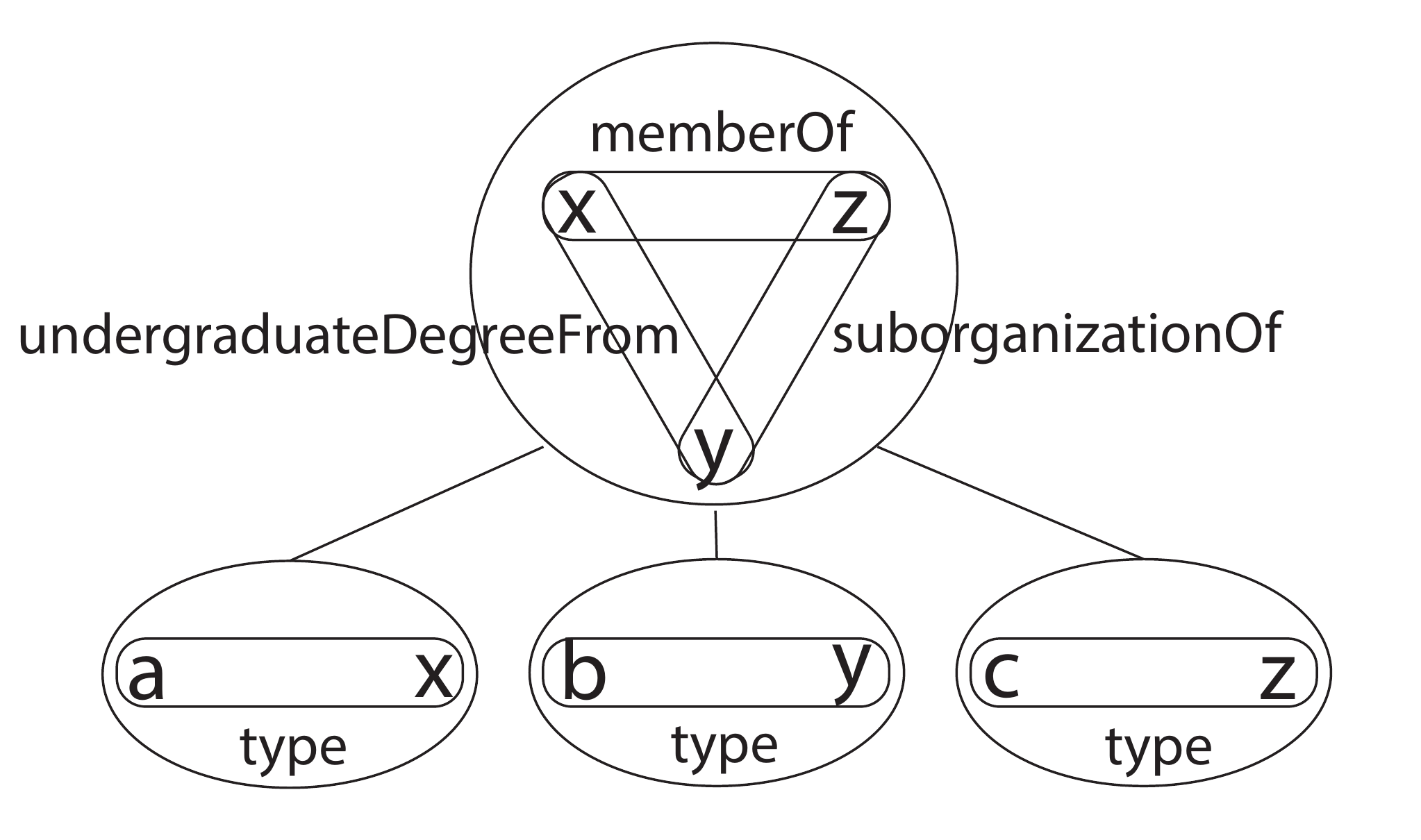}
  \caption{GHD for LUBM query 2.}
  \label{fig:ghd2}
\end{figure}

Using GHDs, we can define a non-trivial cardinality estimate based on the sizes of the relations. Define $Q_t$ as the query formed by joining the relations in $\lambda(t)$. The {\bf {\em (fractional) width}} of a GHD is $\mathsf{AGM}(Q_t)$, which is an upper bound on the number of tuples returned by $Q_t$. The {\bf {\em fractional hypertree width (fhw)}} of a hypergraph $H$ is the minimum width of all GHDs of $H$. \EH{} chooses the GHD with the lowest fhw and smallest height by enumerating all possible GHDs. For example, the GHD that is chosen for query 2 of the LUBM benchmark is shown in \Cref{fig:ghd2} and has a fhw of 1.5. 

After choosing a GHD, \EH{} needs to also choose a global attribute order. This determines the order of the attributes in each trie as well as the order in which attributes are processed in the generic worst-case optimal join algorithm. We choose the global attribute order by doing a breadth-first traversal of the GHD: attributes seen earlier in the traversal are earlier in the order. We also apply some heuristics, described in \Cref{sec:attribute_ordering}, to better order high selectivity attributes.

Finally, during execution we perform two passes over the GHD. First, the GHD is traversed in a bottom-up fashion where \Cref{fig:worst_case} executes over each node in the GHD and children pass intermediate results up to their parents. Second, if necessary, we traverse the GHD in a top-down fashion where we use a message passing algorithm \cite{yannakakis} to materialize the final result.

\section{Classic Optimizations}
\label{sec:optimizations}

The LUBM benchmark contains several queries with acyclic join patterns and equality selections with high selectivity, a class of queries that was not considered in the original \EH{} design. As such, we found it necessary to add three classic query optimization techniques to our worst-case optimal design to enable high performance on this class of queries: (1) optimized index layouts, (2) pushing down selections, and (3) pipelining results in our query plan. We describe briefly how each optimization maps to worst-case optimal join algorithms and demonstrate that these optimizations can have over a two orders of magnitude performance improvement on the overall query execution time.

\subsection{Index Layout}
The impact of various index layouts on predicate selections is a well-researched and empirically solved problem in classic join optimizers \cite{ailamaki2002data,abadi2009column}. It is relatively straightforward to apply similar optimizations to the design in \EH{}. Recall that in \EH{} a single trie is analogous to a single index in a standard database. Therefore, \EH{} performs an equality selection by checking whether a set in the trie contains a value. For the bitset layout we can do this lookup in constant time. For the unsigned integer layout, this is done in $O(\log{n})$ using a binary search. As shown in the \texttt{+Layout} column in \Cref{table:speedup}, the mixed layouts can provide up to a 8.22x performance increase over solely the unsigned integer layout on simple queries with equality selections. Correspondingly, we would like the attributes with equality selections to appear first in the trie, since the first level of each trie is usually dense and therefore best represented as a bitset. Next, we describe how we ensure this by reordering attributes within GHD nodes to push down selections.

\begin{table}
	\small
  \begin{center}
      \setlength{\tabcolsep}{3pt}
    \begin{tabular}{@{}lrrrr@{}}
    \toprule
    Query &+Layout &+Attribute &+GHD &+Pipelining\\
    \midrule
	1	&2.10x	&129.85x	&- &-\\
	2	&8.22x	&1.03x		&- &-\\
	4	&2.02x	&12.88x		&69.94x &- \\
	7	&4.35x	&95.01x		&-  &-\\
	8	&2.24x	&1.99x 		&1.5x		&4.67x \\
	14	&7.92x	&234.49x	&- &-\\
    \bottomrule
    \end{tabular}
    \caption{Relative speedup of each optimization on selected LUBM queries with 133 million triples. \texttt{+Layout} refers to \EH{} when using multiple layouts versus solely a unsigned integer array (index layout). \texttt{+Attribute} refers to reordering attributes with selections within a GHD node. \texttt{+GHD} refers to pushing down selections across GHD nodes in our query plan. \texttt{+Pipelining} refers to pipelining intermediate results in a given query plan. ``-'' means the optimization has no effect on the given query.}
    \label{table:speedup}
  \end{center}
\end{table}

\subsection{Pushing Down Selections}

A classic database optimization is to force high selectivity operations to be processed as early as possible in a query plan~\cite{jarke1984query}. In \EH{} we can do this at two different granularities in our query plans: within GHD nodes and across GHD nodes.\footnote{Recall \EH{} executes a GHD query plan in two phases: (1) the generic worst-case optimal join algorithm runs inside of each node in the GHD and (2) the final result is computed by passing intermediate results across nodes. The phases directly correspond to the two granularities at which we push down selections.}

\subsubsection{Within a Node}
\label{sec:attribute_ordering}
In \EH{} pushing down selections within a GHD node corresponds to rearranging the attribute order for the generic worst-case optimal join algorithm that we described in \Cref{sec:backgound}. Recall, the attribute order determines both the order that attributes are processed in \Cref{fig:worst_case} and the order in which the attributes will appear in the trie. 
\vspace{5.5mm}
\begin{example}
Consider LUBM query 14 where the relation \texttt{R} contains the attributes \texttt{(a,x)}:

{
\footnotesize
 \texttt{\textbf{select} x \textbf{from} R \textbf{where} a = `University'}
}

For this trivial query we produce a single node GHD containing attributes \texttt{(a,x)}.\footnote{Note: for the remainder of the paper \texttt{()} denotes an unordered set of attributes while \texttt{[]} denotes an ordered list of attributes.} A attribute ordering of \texttt{[x,a]} in this node means that \texttt{`x'} is the first level of the trie and \texttt{`a'} is the second level of the trie. Thus,  \EH{} would execute this query by probing the second level of the trie for each \texttt{`x'} attribute to determine if there was a corresponding \texttt{`a'} value of \texttt{`University'}. This is much less efficient than selecting the attribute ordering of \texttt{[a,x]}, where \EH{} can simply perform a lookup in the first level of the trie (to find if an \texttt{`a'} value of \texttt{`University'} exists) and, if successful, return the corresponding second level as the result. 
\end{example}

 This same optimization holds for more complex queries, such as LUBM query 2 (\Cref{fig:ghd2}), where the attribute ordering we select is \texttt{[a, b, c, x, y, z]}. We show in the \texttt{+Attribute} column of \Cref{table:speedup} that forcing the attributes with selections or small initial cardinalities to come first can enable an up to 234.49x performance increase.

\begin{figure}
  \centering
  \includegraphics[width=1.0\linewidth]{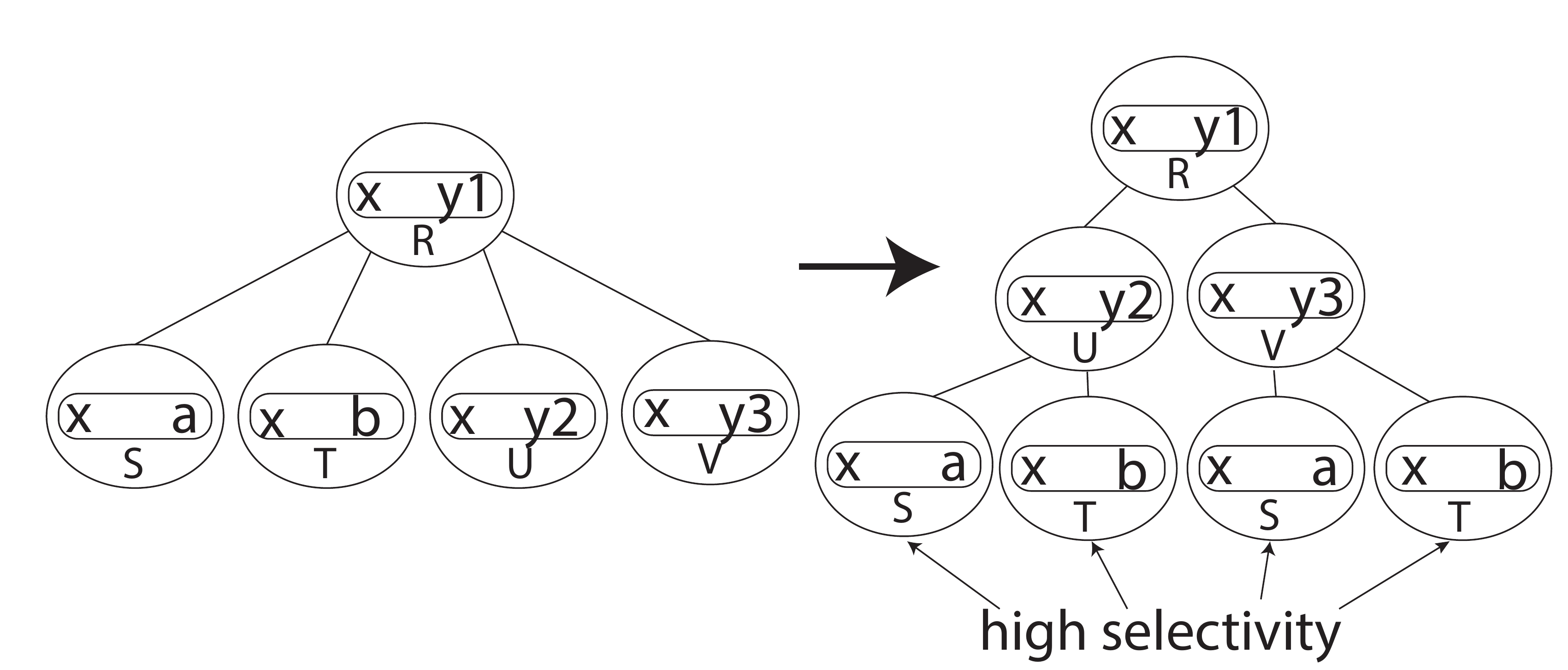}
  \caption{``Across nodes'' transformation of the GHD of LUBM query 4.}
  \label{fig:ghd4}
\end{figure}

\subsubsection{Across Nodes}
Pushing down selections across nodes in \EH's query plans corresponds to 
changing the criteria for choosing a GHD described in \Cref{sec:backgound}. Our goal is to have high selectivity or low-cardinality nodes be pushed down far as possible in the GHD (so that they are executed earlier in our bottom-up pass). We accomplish this by adding three additional steps to our GHD optimizer:

\begin{enumerate}
\item Find optimal GHDs $\mathcal{T}$ with respect to fhw, changing $V$ in the AGM constraint to be only the attributes without selections.
\item Let $R_s$ be some relations with selections and let $R_t$ be the relations that we plan to place in a subtree. If for each $e \in R_s$, there exists $e' \in R_t$ such that $e'$ covers $e$'s unselected attributes, include $R_s$ in the subtree for $R_t$. This means that we may duplicate some members of $R_s$ to include them in multiple subtrees. 
\item Of the GHDs $\mathcal{T}$, choose a $T \in \mathcal{T}$ with maximal selection depth, where selection depth is the sum of the distances from selections to the root of the GHD. 
\end{enumerate}

\begin{example}
Consider the acyclic join pattern for LUBM query 4 containing high selectivity attributes \texttt{(a,b)}:
{
\footnotesize
$$\texttt{R(x,y1)} \bowtie \texttt{S(x,a)} \bowtie \texttt{T(x,b)} 
 \bowtie \texttt{U(x,y2)} \bowtie \texttt{V(x,y3)} $$
}
\Cref{fig:ghd4} shows two possible GHDs for this query. The GHD on the left is the one produced without using the three steps above. This GHD does not filter out any intermediate results across potentially high selectivity nodes when results are first passed up the GHD. The GHD on the right uses the three steps above. Here the nodes with attributes \texttt{`a'} and \texttt{`b'} are below all other nodes in the GHD, ensuring that high selectivity attributes are processed early in the query plan. 
\end{example}

This optimization applies to only two queries in the LUBM benchmark, but provides up to a 69.94x speedup as we show in the \texttt{+GHD} column of \Cref{table:speedup}. 

\subsection{Pipelining}
Finally, pipelining is a classic query optimization used to reduce the size of materialized intermediate results in a query plan \cite{jarke1984query}. We add a simple rule such that the root node of a GHD can be pipelined with one child node in \EH{}:

\begin{definition}
Given a GHD $T$, we say $(t_0,t_1) \in V(T) \times V(T) $ are {\em pipelineable} if $t_0 \neq t_1$ and $\chi(t_0) \cap
\chi(t_1)$ is a prefix of the trie orders for both $t_0$ and $t_1$.
\label{def:pipeline}
\end{definition}

\begin{example}
Consider the following query pattern from LUBM query 8 over relation \texttt{R} with attributes \texttt{(x,y)} and relation \texttt{S} with attributes \texttt{(x,z)}:

{
\footnotesize
 \texttt{\textbf{select} x,y,z \textbf{from} R,S \textbf{where} R.x = S.x}
}

The GHD \EH{} produces for this query contains two nodes with respective ordered attributes \texttt{[x,y]} (root $t_0$) and \texttt{[x,z]} (child $t_1$). By definition this GHD is pipelineable as the nodes share the common prefix \texttt{`x'}.
\end{example}

On LUBM query 8 we found that pipelining results between nodes with materialized attributes provided up to a 4.67x performance advantage as shown in the \texttt{+Pipelining} column of \Cref{table:speedup}. Unfortunately, the impact of pipelining is negligible on most LUBM queries as the number of output attributes is often small and so are the intermediate cardinalities.

\section{Experiments}
\label{sec:experiments}

We benchmark a standard relational engine, two worst-case optimal join engines, and two state-of-the-art specialized RDF engines on the LUBM RDF benchmark.  We select MonetDB as the classical relational data processing engine baseline, LogicBlox and \EH{} as the worst-case optimal engine baselines, and RDF-3X and TripleBit as the specialized RDF engine baselines. Our comparison shows that \EH{} and LogicBlox's designs outperform all other engines on cyclic queries, where, again, pairwise joins are suboptimal. On the remaining queries we show how \EH{} remains competitive with the specialized RDF engines due to our addition of the three classical database optimizations described in \Cref{sec:optimizations}.

\subsection{Experiment Setup}
We describe the details of our experimental setting.

\subsubsection{LUBM Benchmark}
The LUBM benchmark is a standard RDF benchmark with a synthetic data generator \cite{guo2005lubm}. The data generator produces RDF data representing a university system ontology. We generated 133 million triples for the comparisons in this section. The LUBM benchmark contains complex multiway star join patterns as well as two cyclic queries with triangle patterns. We run the complete LUBM benchmark while removing the inference step for each query. This is standard in benchmarking comparisons \cite{atre2010matrix,yuan2013triplebit}.  We omit queries 6 and 10, since without the inference step, they correspond to other queries in the benchmark. 

\subsubsection{Comparison Engines}

We describe the specialized RDF engines (TripleBit and RDF-3X) and general purpose relational engines (MonetDB and LogicBlox) with which we compare.

\paragraph*{RDF Engines}
We compare against RDF-3X v0.3.8 and TripleBit, two high performance shared memory RDF engines.  TripleBit \cite{yuan2013triplebit} and RDF-3X \cite{neumann2010rdf} have been shown to consistently outperform traditional column and row store databases \cite{yuan2013triplebit,neumann2010rdf}.  RDF-3X is a popular and established RDF engine which performs well across a variety of SPARQL queries. RDF-3X builds a full set of permutations on all triples and uses selectivity estimates to choose the best join order. TripleBit \cite{yuan2013triplebit} is a more recent RDF engine which uses a sophisticated matrix representation and has been shown to compete with and often outperform RDF-3X on a range of RDF queries on larger scale data. TripleBit reduces the size of the data and indexes through two auxiliary data structures to minimize the cost of index selection during query evaluation. Both engines use optimizers that generate optimal join orderings.

\paragraph*{Relational Engines}
We also provide comparisons to MonetDB (Jul2015-SP1 release) and LogicBlox v4.3.4 which are two general purpose relational engines. MonetDB is a popular open source column store database whose performance has been shown to outperform row store designs, such as PostgreSQL, by orders of magnitude on RDF workloads \cite{neumann2010rdf}. LogicBlox is a commercial database engine that uses a worst-case optimal join algorithm similar to that inside of the \EH{} engine. For all relational engines, including \EH{}, we store and process the RDF data in a vertically partitioned manner as this has been shown to be superior to storing the data as triples \cite{abadi2007scalable,neumann2010rdf}. Vertical partitioning is the process of grouping the triples by their predicate name, with all triples sharing the same predicate name being stored under a table denoted by the predicate name \cite{abadi2007scalable}.

\subsubsection{Experimental Setting}
We ran all experiments on a single machine with a total of 48 cores on four Intel Xeon E5-4657L v2 CPUs and 1 TB of RAM. We compiled the C++ engines (RDF-3X, TripleBit, \EH{}) with g++ 4.9.3 (-O3). For all engines, we chose buffer sizes and heap sizes that were at least an order of magnitude larger than the dataset itself to avoid garbage collection. For all engines, we put the database files in \emph{tmpfs}, a RAM disk, which is a standard resource to use when comparing in-memory engines to databases \cite{kim2015taming}. For MonetDB, we explicitly built indexes on each column and each pair of columns as well as histograms over each relation (with the \texttt{ANALYZE} command).

\subsubsection{Metrics}
For each query we measure the wall clock time of the engine to complete the query. We do not measure the time for data loading, index creation, or query output for each engine. For TripleBit and RDF-3X, we do not include the time for lookups from ID to String (or output time) at the end of the query. We run each query seven times, discarding the worst and best runtimes while reporting the average of the remaining times. We do not measure compilation time for \EH{}. Since we run queries back-to-back, often only the first execution incurs compilation costs, and this longest run is discarded for all engines.

\subsection{End-to-End Comparison}

\begin{table}
	\small
  \begin{center}
      \setlength{\tabcolsep}{3pt}
    \begin{tabular}{@{}lrrrrrrrrr@{}}
    \toprule
    Query &Best  & EH                  &TripleBit         	  &RDF-3X    	 &MonetDB &LogicBlox \\
    \midrule
	Q1 &\multicolumn{1}{r|}{4.00}	&1.51x	&3.45x	&\textbf{1.00x}	&174.58x	&8.62x \\
	Q2 &\multicolumn{1}{r|}{973.95}	&\textbf{1.00x}	&2.38x	&1.92x	&8.79x	&1.52x \\
	Q3 &\multicolumn{1}{r|}{0.47}	&\textbf{1.00x}	&92.61x	&8.44x	&283.37x	&83.41x \\
	Q4 &\multicolumn{1}{r|}{3.39}	&4.62x	&\textbf{1.00x}	&1.77x	&2093.78x	&116.32x \\
	Q5 &\multicolumn{1}{r|}{0.44}	&\textbf{1.00x}	&99.21x	&9.15x	&303.11x	&81.44x \\
	Q7 &\multicolumn{1}{r|}{6.00}	&3.18x	&8.53x	&\textbf{1.00x}	&573.33x	&6.52x \\
	Q8 &\multicolumn{1}{r|}{78.50}	&9.83x	&\textbf{1.00x}	&3.07x	&206.62x	&5.03x \\
	Q9 &\multicolumn{1}{r|}{581.37}	&\textbf{1.00x}	&3.53x	&6.63x	&24.29x	&1.35x \\
	Q11 &\multicolumn{1}{r|}{0.45}	&\textbf{1.00x}	&6.07x	&11.03x	&58.63x	&73.76x \\
	Q12 &\multicolumn{1}{r|}{3.05}	&2.22x	&\textbf{1.00x}	&7.86x	&118.94x	&50.23x \\
	Q13 &\multicolumn{1}{r|}{0.87}	&\textbf{1.00x}	&48.90x	&35.49x	&86.18x	&102.77x \\
	Q14 &\multicolumn{1}{r|}{3.00}	&1.90x	&54.47x	&\textbf{1.00x}	&313.47x	&325.02x \\
    \bottomrule
    \end{tabular}
    \caption{Runtime in milliseconds for best performing system and relative runtime for each engine on the LUBM benchmark with 133 million triples. EH denotes the \EH{} engine.}
    \label{table:rdf}
  \end{center}
\end{table}

LUBM queries 2 and 9 are the two cyclic queries that contain a triangle pattern. Unsurprisingly, here LogicBlox outperforms specialized engines by 3-5x and MonetDB by 17.96x (\Cref{table:rdf}) due to the asymptotic advantage of worst-case optimal join algorithms. On these queries \EH{} is 1.5x faster than LogicBlox due to our set layouts, which are designed for single-instruction multiple data parallelism \cite{aberger2015emptyheaded}. In general, our speedup over LogicBlox is more modest here than on previously reported cyclic graph patterns \cite{aberger2015emptyheaded} due to the presence of high selectivity selections. 

On acyclic queries with high selectivity, \EH{} also competes with the specialized RDF engines. On simple acyclic queries with selections (LUBM 1,3,5,11,13,14), \EH{} is able to provide covering indexes, like the specialized engines, using only our trie data structure and the attribute order we described in \Cref{sec:attribute_ordering}. Therefore \EH{} maintains competitive performance with RDF-3X and TripleBit (\Cref{table:rdf}).  On more complex acyclic queries with selections (LUBM 7,8,12), RDF-3X and TripleBit observe a performance advantage over \EH{} due to their sophisticated cost-based query optimizers which combine selectivity estimates and join order (\Cref{table:rdf}). Our optimizations from \Cref{sec:optimizations} can provide up to a 48x performance improvement here, but more sophisticated optimizations are needed to outperform the specialized engines. Finally, on LUBM query 8 we observe a performance slowdown when compared to LogicBlox. This is due to expensive reallocations that occur within the \EH{} engine. When removing allocations, we observed that \EH's performance for query 8 was equivalent to that of RDF-3X. We hope to add these further optimizations to the \EH{} engine in the future.

\section{Conclusion}
\label{sec:conclusion}
Our work shows that worst-case optimal join algorithms can provide up to a 6x performance advantage on bottleneck cyclic RDF queries when compared to the join algorithms used inside of both specialized RDF engines and traditional databases. On simple acyclic queries, we show how three classic database optimizations can be added to a generic worst-case optimal join processing engine to enable sufficient performance on simple acyclic RDF queries. We hope our work serves as a feasibility study, validating that worst-case optimal join algorithms can benefit common RDF workloads.

{
\scriptsize
\noindent\textbf{Acknowledgments}
We thank the authors of the LogicBlox engine for their assistance in our experimental comparisons. We gratefully acknowledge the support of the Defense 
Advanced Research Projects Agency (DARPA) XDATA Program 
under No. FA8750-12-2-0335 and DEFT Program under 
No. FA8750-13-2-0039, DARPA's MEMEX program and SIMPLEX 
program, the National Science Foundation (NSF) CAREER 
Award under No. IIS-1353606, the Office of Naval Research 
(ONR) under awards No. N000141210041 and No. N000141310129,
and Intel. Any opinions, 
findings, and conclusions or recommendations expressed in this 
material are those of the authors and do not necessarily 
reflect the views of DARPA, AFRL, NSF, ONR, or the 
U.S. government. 
}

\bibliographystyle{plain}
\bibliography{references}

\begin{thebibliography}{10}

\bibitem{abadi2009column}
Daniel~J Abadi, Peter~A Boncz, and Stavros Harizopoulos.
\newblock Column-oriented database systems.
\newblock {\em VLDB '09'}, pages 1664--1665.

\bibitem{abadi2007scalable}
Abadi et~al.
\newblock Scalable semantic web data management using vertical partitioning.
\newblock In {\em VLDB `07}, pages 411--422. VLDB Endowment, 2007.

\bibitem{aberger2015emptyheaded}
Aberger et~al.
\newblock Emptyheaded: A relational engine for graph processing.
\newblock {\em arXiv preprint arXiv:1503.02368}, 2015.

\bibitem{ailamaki2002data}
Ailamaki et~al.
\newblock Data page layouts for relational databases on deep memory
  hierarchies.
\newblock {\em VLDB `02}, 11(3):198--215.

\bibitem{aref2015design}
Aref et~al.
\newblock Design and implementation of the logicblox system.
\newblock In {\em SIGMOD `15}, pages 1371--1382. ACM, 2015.

\bibitem{astrahan1976system}
Astrahan et~al.
\newblock System r: relational approach to database management.
\newblock {\em TODS}, 1(2):97--137, 1976.

\bibitem{atre2010matrix}
Atre et~al.
\newblock Matrix bit loaded: a scalable lightweight join query processor for
  rdf data.
\newblock In {\em WWW '10}, pages 41--50. ACM.

\bibitem{agm}
Atserias et~al.
\newblock Size bounds and query plans for relational joins.
\newblock {\em SIAM Journal on Computing}, 42(4):1737--1767, 2013.

\bibitem{green2007provenance}
Green et~al.
\newblock Provenance semirings.
\newblock In {\em SIGMOD '07'}, pages 31--40.

\bibitem{guo2005lubm}
Guo et~al.
\newblock Lubm: A benchmark for owl knowledge base systems.
\newblock {\em Web Semantics: Science, Services and Agents on the World Wide
  Web}, 3(2):158--182, 2005.

\bibitem{joglekar2015aggregations}
Joglekar et~al.
\newblock Aggregations over generalized hypertree decompositions.
\newblock {\em arXiv preprint arXiv:1508.07532}, 2015.

\bibitem{kim2015taming}
Kim et~al.
\newblock Taming subgraph isomorphism for rdf query processing.
\newblock {\em VLDB `15}, 8(11):1238--1249, 2015.

\bibitem{klyne2006resource}
Klyne et~al.
\newblock Resource description framework (rdf): Concepts and abstract syntax.
\newblock 2006.

\bibitem{neumann2010rdf}
Neumann et~al.
\newblock The rdf-3x engine for scalable management of rdf data.
\newblock {\em VLDB `10}, 19(1):91--113.

\bibitem{ngo2012worst}
Ngo et~al.
\newblock Worst-case optimal join algorithms:[extended abstract].
\newblock In {\em PODS}, pages 37--48. ACM, 2012.

\bibitem{Hong:2012:GDE:2150976.2151013}
S.~Hong et~al.
\newblock Green-marl: A dsl for easy and efficient graph analysis.
\newblock ASPLOS XVII, pages 349--362, 2012.

\bibitem{stonebraker2005c}
Stonebraker et~al.
\newblock C-store: a column-oriented dbms.
\newblock In {\em VLDB `05}, pages 553--564. VLDB Endowment, 2005.

\bibitem{yuan2013triplebit}
Yuan et~al.
\newblock Triplebit: a fast and compact system for large scale rdf data.
\newblock {\em VLDB `13}, 6(7):517--528.

\bibitem{jarke1984query}
Matthias Jarke and Jurgen Koch.
\newblock Query optimization in database systems.
\newblock {\em ACM Computing surveys (CsUR)}, 16(2):111--152, 1984.

\bibitem{yannakakis}
Mihalis Yannakakis.
\newblock Algorithms for acyclic database schemes.
\newblock In {\em VLDB}, pages 82--94, 1981.

\end{thebibliography}

\begin{appendix}

\subsection{Related Work}

Our work extends previous research on three distinct types of data processing engines: specialized RDF engines, traditional relational engines, and worst-case optimal engines.

\paragraph*{Pairwise Engines}
The most straightforward way to store RDF data in a traditional general purpose RDBMS is to store all Subject-Predicate-Object triples in a 3-column table, called the triple table \cite{neumann2010rdf}, where each row holds a triple. Unfortunately, querying a triple table with millions of rows is rarely optimal. For example, many RDF queries involve self-joins and a huge triple table complicates both selectivity estimations and increases the time for simple operations such as scans \cite{abadi2007scalable,yuan2013triplebit}. Another approach, vertical partitioning, stores RDF data in many two-column tables, one for each unique predicate \cite{abadi2007scalable}. Like storing a triple table, this approach can be used in either row-oriented or column-oriented databases. In this paper we show for the first time that it can be naturally mapped to worst-case optimal databases as well. 

\paragraph*{RDF Engines} Two of the most popular specialized RDF engines are RDF-3X and TripleBit. Both accept queries in the SPARQL query language and have been shown to significantly outperform traditional relational engines. RDF-3X creates a full set of subject-predicate-object indexes by building clustering B+ trees on all six permutations of the triples \cite{neumann2010rdf}. RDF-3X also maintains nine aggregate indexes, which include all six binary and all three unary projections. Each index provides some selectivity estimates and the aggregate indexes are used to select the fastest index for a given query. In the TripleBit engine, RDF triples are represented using a compact matrix representation \cite{yuan2013triplebit}. TripleBit also stores two auxiliary index structures and two binary aggregate indexes to use the selectivity estimation of query patterns to select the most effective indexes, minimize the number of indexes needed, and determine the query plan. Like \EH, both RDF-3X and TripleBit use dictionary encoding.

\paragraph*{Multi-Way Engines}
The first worst-case optimal join algorithm was recently derived \cite{ngo2012worst}. The LogicBlox (LB) engine \cite{aref2015design} is the first commercial database engine to use a worst-case optimal algorithm. Recent theoretical advances \cite{joglekar2015aggregations} have suggested worst-case optimal join processing is applicable beyond standard join pattern queries. We continue in this line of work, applying worst-case optimal algorithms to a standard RDF workload.

\subsection{LUBM Queries}

We provide the SPARQL syntax used for each query run in this paper as well as the output cardinality of each query run with 133 million triples produced by the LUBM data generator.

All queries include the following prefixes:

{
\scriptsize
\begin{lstlisting}
PREFIX rdf: 
 <http://www.w3.org/1999/02/22-rdf-syntax-ns#>
PREFIX ub: 
 <http://www.lehigh.edu/~zhp2/2004/0401/univ-bench.owl#>
\end{lstlisting}
}
\textbf{Query 1:} 4 tuples
{
\scriptsize
\begin{lstlisting}
SELECT ?X	
WHERE{
  ?X rdf:type ub:GraduateStudent .
  ?X ub:takesCourse 
  <http://www.Department0.University0.edu/GraduateCourse0>}
\end{lstlisting}
}

\textbf{Query 2:} 2,528 tuples
{
\scriptsize
\begin{lstlisting}
SELECT ?X ?Y ?Z
WHERE{
  ?X rdf:type ub:GraduateStudent .
  ?Y rdf:type ub:University .
  ?Z rdf:type ub:Department .
  ?X ub:memberOf ?Z .
  ?Z ub:subOrganizationOf ?Y .
  ?X ub:undergraduateDegreeFrom ?Y}
\end{lstlisting}
}

\textbf{Query 3:} 6 tuples
{
\scriptsize
\begin{lstlisting}
SELECT ?X
WHERE{
 ?X rdf:type ub:Publication .
 ?X ub:publicationAuthor 
  <http://www.Department0.University0.edu/
     AssistantProfessor0>}
\end{lstlisting}
}

\textbf{Query 4:} 14 tuples
{
\scriptsize
\begin{lstlisting}
SELECT ?X ?Y1 ?Y2 ?Y3
WHERE{
  ?X rdf:type ub:AssociateProfessor .
  ?X ub:worksFor <http://www.Department0.University0.edu> .
  ?X ub:name ?Y1 .
  ?X ub:emailAddress ?Y2 .
  ?X ub:telephone ?Y3}
\end{lstlisting}
}

\textbf{Query 5:} 532 tuples
{
\scriptsize
\begin{lstlisting}
SELECT ?X
WHERE{ 
  ?X rdf:type ub:UndergraduateStudent .
  ?X ub:memberOf <http://www.Department0.University0.edu>}
\end{lstlisting}
}

\textbf{Query 7:} 59 tuples
{
\scriptsize
\begin{lstlisting}
SELECT ?X ?Y
WHERE{
  ?X rdf:type ub:UndergraduateStudent .
  ?Y rdf:type ub:Course .
  ?X ub:takesCourse ?Y .
  <http://www.Department0.University0.edu/
     AssociateProfessor0> ub:teacherOf ?Y}
\end{lstlisting}
}

\textbf{Query 8:} 5,916 tuples
{
\scriptsize
\begin{lstlisting}
SELECT ?X ?Y ?Z
WHERE{
  ?X rdf:type ub:UndergraduateStudent .
  ?Y rdf:type ub:Department .
  ?X ub:memberOf ?Y .
  ?Y ub:subOrganizationOf <http://www.University0.edu> .
  ?X ub:emailAddress ?Z}
\end{lstlisting}
}
\textbf{Query 9:} 44,021 tuples
{
\scriptsize
\begin{lstlisting}
SELECT ?X ?Y ?Z
WHERE{
  ?X rdf:type ub:UndergraduateStudent .
  ?Y rdf:type ub:Course .
  ?Z rdf:type ub:AssistantProfessor .
  ?X ub:advisor ?Z .
  ?Z ub:teacherOf ?Y .
  ?X ub:takesCourse ?Y}
\end{lstlisting}
}

\textbf{Query 11:} 0 tuples
{
\scriptsize
\begin{lstlisting}
SELECT ?X
WHERE{
  ?X rdf:type ub:ResearchGroup .
  ?X ub:subOrganizationOf <http://www.University0.edu>}
\end{lstlisting}
}

\vspace{20mm}

\textbf{Query 12:} 125 tuples
{
\scriptsize
\begin{lstlisting}
SELECT ?X ?Y
WHERE{
  ?X rdf:type ub:FullProfessor .
  ?Y rdf:type ub:Department .
  ?X ub:worksFor ?Y .
  ?Y ub:subOrganizationOf <http://www.University0.edu>}
\end{lstlisting}
}

\textbf{Query 13:} 2,489 tuples
{
\scriptsize
\begin{lstlisting}
SELECT ?X
WHERE{
  ?X rdf:type ub:GraduateStudent .
  ?X ub:undergraduateDegreeFrom <http://www.University567.edu>}
\end{lstlisting}
}

\textbf{Query 14:} 7,924,765 tuples
{
\scriptsize
\begin{lstlisting}
SELECT ?X
WHERE {?X rdf:type ub:UndergraduateStudent}
\end{lstlisting}
}

\end{appendix}

\end{document}